\documentclass[aps,prl,twocolumn,superscriptaddress]{revtex4}
\usepackage{graphicx}
\usepackage[latin1]{inputenc}
\usepackage{textcomp}
\usepackage{mathptmx}
\begin{document}

\title{Observation of Stable Jones-Roberts Solitons in Bose-Einstein Condensates}

\author{Nadine Meyer}
\affiliation{Midlands Ultracold Atom Research Centre, School of Physics and Astronomy, University of Birmingham, Edgbaston, Birmingham, B15 2TT, United Kingdom}
\affiliation{ICFO-Institut de Ciencies Fotoniques, The Barcelona Institute of Science and Technology, 08860 Castelldefels, Barcelona, Spain}

\author{Harry Proud}
\affiliation{Midlands Ultracold Atom Research Centre, School Of Physics and Astronomy, University of Birmingham, Edgbaston, Birmingham, B15 2TT, United Kingdom}
\author{Marisa Perea-Ortiz}
\affiliation{Midlands Ultracold Atom Research Centre, School Of Physics and Astronomy, University of Birmingham, Edgbaston, Birmingham, B15 2TT, United Kingdom}
\author{Charlotte O'Neale}
\affiliation{Midlands Ultracold Atom Research Centre, School Of Physics and Astronomy, University of Birmingham, Edgbaston, Birmingham, B15 2TT, United Kingdom}
\affiliation{IOP Publishing, Temple Way, Bristol BS1 6HG, United Kingdom}
\author{Mathis Baumert}
\affiliation{Midlands Ultracold Atom Research Centre, School Of Physics and Astronomy, University of Birmingham, Edgbaston, Birmingham, B15 2TT, United Kingdom}
\affiliation{Abaco  Systems  Limited,  Tove  Valley  Business  Park,  Towcester,  Northamptonshire,  NN12  6PF, United Kingdom}
\author{Michael Holynski} 
\affiliation{Midlands Ultracold Atom Research Centre, School Of Physics and Astronomy, University of Birmingham, Edgbaston, Birmingham, B15 2TT, United Kingdom}
\author{Jochen Kronj\"ager}
\affiliation{Midlands Ultracold Atom Research Centre, School Of Physics and Astronomy, University of Birmingham, Edgbaston, Birmingham, B15 2TT, United Kingdom}
\affiliation{National Physics Laboratory, Hampton Road, Teddington, Middlesex, TW11 0LW, United Kingdom}
\author{Giovanni Barontini}
\email{g.barontini@bham.ac.uk}
\author{Kai Bongs}
\affiliation{Midlands Ultracold Atom Research Centre, School Of Physics and Astronomy, University of Birmingham, Edgbaston, Birmingham, B15 2TT, United Kingdom}

\date{\today}

\begin{abstract}
We experimentally generate two-dimensional Jones-Roberts solitons in a three-dimensional atomic Bose-Einstein condensate by imprinting a triangular phase pattern. By monitoring their dynamics we observe that this kind of solitary waves are resistant to both dynamic (snaking) and thermodynamic instabilities, that usually are known to strongly limit the lifetime of dark plane solitons in dimensions higher than one. We additionally find signatures of a possible dipole-like interaction between them. Our results confirm that Jones-Roberts solitons are stable solutions of the non-linear Schr\"odinger equation in higher dimensions and promote these excitations for applications beyond matter wave physics, like energy and information transport in noisy and inhomogeneous environments. 
\end{abstract}


\maketitle 

Waves play a key role in physics and technology, ranging from quantum mechanics to telecommunications. In linear media, waves spread both transversally and longitudinally due to dispersion, making them unsuitable for directed transport. This is at variance of non-linear media, where solitary waves (solitons) become possible. In solitons, the broadening due to dispersion is counteracted by a non-linear compression, leading to form-stable propagation at subsonic speed over large distances and particle-like properties. Solitons in water channels have first been reported in 1844 \cite{SR} and have since been found in as diverse areas as high-speed data communication in optical fibres \cite{Kiv,Kiv2}, energy transport along DNA in biology \cite{Lautrup,Heimburg} and tropospheric phenomena like 'Morning glory' clouds \cite{Christie}. In experiments with ultracold atoms, dark plane solitons (DPS) have been extensively studied \cite{Anderson,Becker,Burger,Dens,Streck}, but in systems with dimensions larger than one a rapid decay due to thermodynamical or dynamical (snaking) instabilities has always been observed \cite{Fed,Mury}. Indeed, so far stable solitons and thus dispersion-free wave transport have been restricted only to one-dimensional systems \cite{Becker}. In this Letter we report the experimental realisation of form-stable excitations in higher dimensions, which have been predicted by Jones and Roberts in 1982 \cite{JR}, but that so far remained elusive. To create these Jones-Roberts solitons (JRS) we imprint a specific phase structure onto an atomic Bose-Einstein condensate (BEC). We find that JRSs are immune to both dynamical and thermodynamical instabilities and indeed their lifetime greatly excesses the one of simple DPSs, created with the same technique. We characterize our JRSs in terms of their size, speed and direction of propagation, finding good agreement with numerical simulations and observing an interesting signature of dipole-like interaction. In addition we demonstrate that our JRSs fulfil the Kadomtsev-Petviashvili condition, as predicted by Jones and Roberts in their seminal work \cite{JR}. 

\begin{figure}
	\centering
		\includegraphics[width=0.5\textwidth]{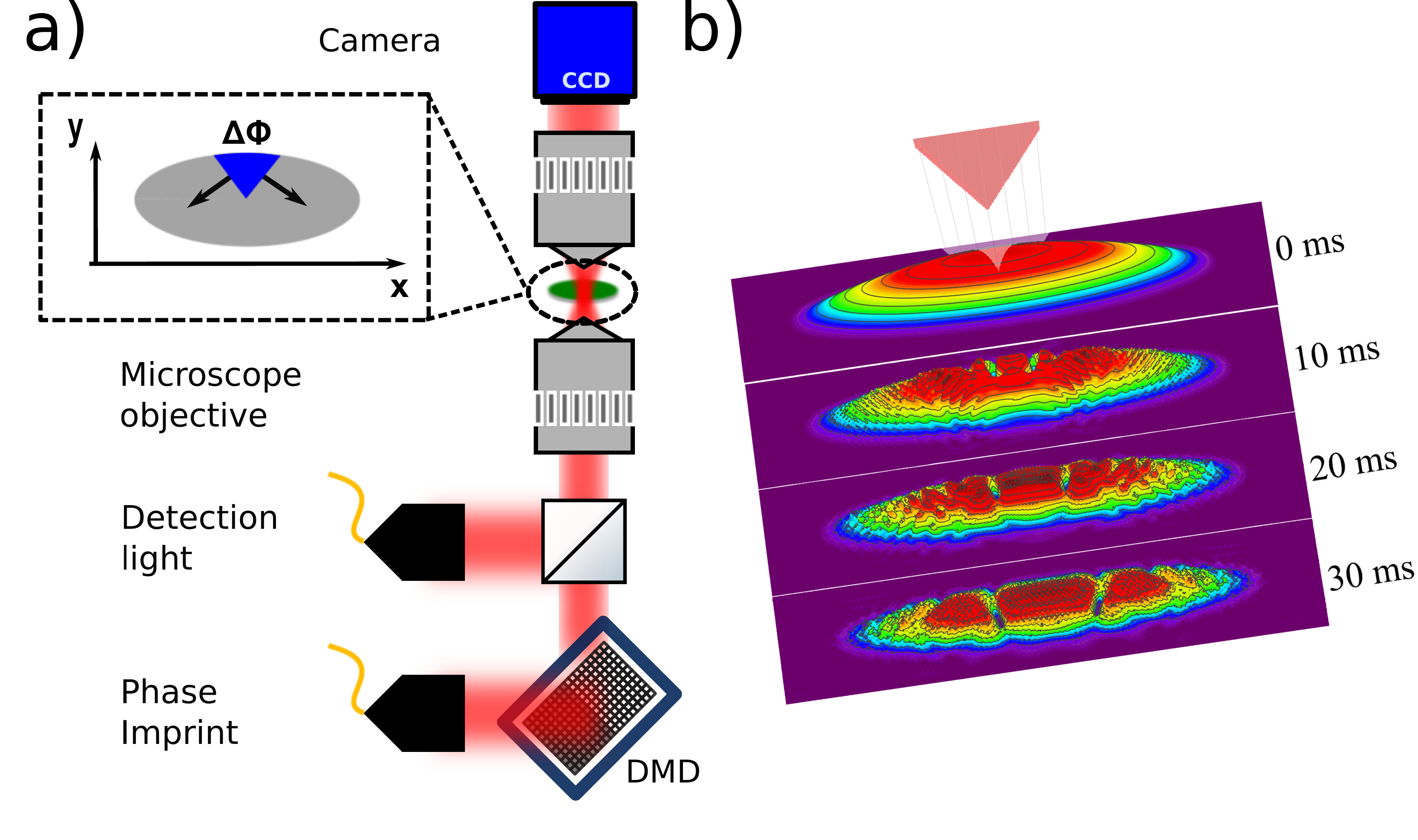}
	\caption{(Color online) a) Experimental setup for phase imprinting. An imprint pattern displayed on a digital micro-mirror device (DMD) is imaged by a microscope objective onto the atoms (see inset). The detection light is superimposed on a polarizing beam splitting cube and transverses the atomic cloud, whose shadow is imaged with a second microscope objective onto the chip of a CCD camera. b) Gross-Pitaevskii simulations of the creation of JRS via phase engineering. A triangular phase pattern whose lower vertex is approximately at the centre of the BEC generates two JRSs that travel across the BEC.}
\label{Iw}
\end{figure}

Within the family of soliton excitations, JRSs have been predicted as the only class of form stable density disturbances in higher dimensions \cite{JR}. In contrast to DPS, JRS are well localized density dips with a finite extent in \emph{all} dimensions.  Their shape and properties depend on their their speed $v$. In two dimensions, for $v$ finite but lower than the speed of sound $c$, they take the form of finite line-shaped density minima, called rarefaction pulses. In the limit $v\rightarrow0$ instead, they transform into spatially separated vortex-antivortex (VA) pairs that mutually propel each other \cite{JR}. In the first case the vorticity vanishes and the phase pattern shows a dipolar structure of two phase winding points of opposite sign connected via an elongated phase step \cite{Smirnov}. In the case of spatially separated VA pairs, the phase pattern consists, as one would expect, of two separated 2$\pi$ phase winding points of opposite charge. A similar picture holds for 3D systems, where JRS take the form of axisymmetric solitary waves transitioning from rarefaction pulses to vortex rings, instead that to VA pairs. Both in two and three dimension, they feature a distinctive elongated shape that allows propagation without change of form and, due to their reduced size and area, they are expected to be immune to the snaking instability and to be resilient against scattering of thermal excitations \cite{Tsuchi}.

In our experiment, we start loading in 10 s $~10^8$ $^{87}$Rb atoms in a 3D MOT fed by a 2D MOT. After an optical molasses stage the atoms are pumped in the $|F=2,m_F=2\rangle$ state and then magnetically transported to the science chamber by moving the trapping coils. In the science chamber the atoms are further cooled by radio frequency evaporation and then transferred into an optical dipole trap. This is produced by a single astigmatic beam at 1550 nm, focussed to a waist of 11 $\mu$m. The distance between the two foci is approximately 1 mm. The atoms are transferred into the region where the beam is focussed vertically, providing a strong confinement on the vertical direction and much weaker confinement in the horizontal plane. A nearly pure BEC of typically 4$\times$10$^4$ atoms in the $|F=2\rangle$ ground state hyperfine manifold is then formed with a subsequent evaporation. The final trapping frequencies are 2$\pi\times$(5,30,250) Hz, leading to an oblate BEC. 

To create the JRSs, we employ a phase imprinting method \cite{Dobrek}. The phase imprinting and imaging setup is illustrated in Fig. 1a. A near-resonant light is reflected by a digital micro-mirror device and then sent onto the atoms along the vertical direction using a high resolution optical microscope objective. Each of the 1920$\times$1080 micro-mirrors can be individually controlled allowing to imprint on the reflected beam any arbitrary intensity pattern $I(x,y)$. The phase of the atoms is therefore locally changed by inducing a dipole potential $U_{dip}(x,y)\propto I(x,y)/\Delta$, where $\Delta$ is the detuning with respect to the atomic transition. The detection light is superimposed to the imprinting beam on a polarising beam cube. The atoms are imaged by absorption imaging with a CCD camera using a second microscope objective that allows a resolution of $\simeq$1 $\mu$m. 

\begin{figure}
	\centering
		\includegraphics[width=0.5\textwidth]{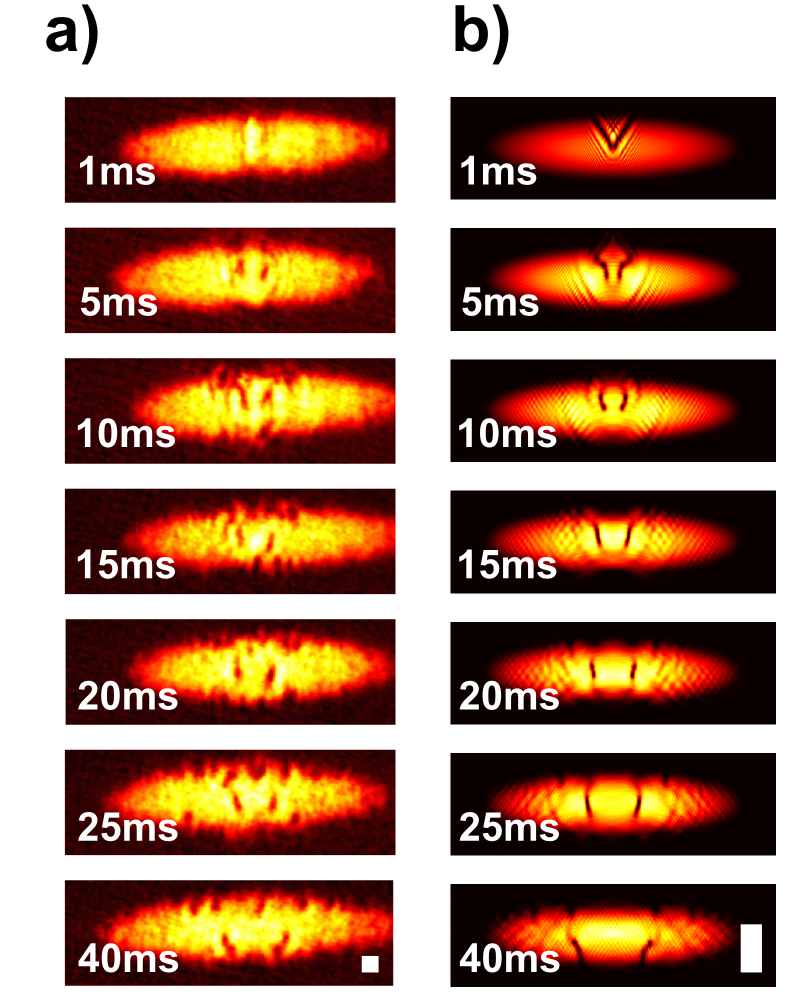}
	\caption{(Color online) a) Experimental density profiles of the BEC taken at different times after the initial imprinting. The time of flight is 10ms. b) Corresponding numerical simulations (in trap) using the Gross-Pitaevskii equation. }
\label{Iw}
\end{figure}

By performing numerical simulations using the Gross-Pitaevskii equation (GPE), we have found that imprinting a homogeneous triangular-shaped phase structure on our BEC leads to the nucleation of a couple of JRS that travel in opposite directions, as shown in Fig. 1b. The triangular shape combines two key features of JRS: phase winding around its vertices and an elongated phase profile along its edges. To identify the most efficient way to create the JRSs we have performed a systematic numerical study changing the shape and the position of the imprinted triangle. We have found that a triangle whose lower vertex subtends an angle of 90 degrees and that imprints a phase difference of $\pi$ is the best choice to create 2 long-living JRSs. Smaller subtended angles lead to the creation of JRSs too close to the upper border of the BEC, limiting their lifetime. Larger angles instead launch the JRSs more in the direction of the short axis, also shortening the time they can travel through the BEC. Finally, we have found that imprinting a phase step lower than 0.9$\pi$ leads to the same effect as imprinting a triangle with a smaller subtended angle. For these reasons, in the experiment the imprinted triangle has the lower vertex positioned at the centre of the BEC and subtends an angle of 90 degrees. We illuminate the BEC with a light pulse at approximately 10 GHz detuning from the $|F=2\rangle\rightarrow|F'=3\rangle$ transition for 28 $\mu$s to create a phase difference of $~\pi$. Before performing absorption imaging on the BEC along the vertical direction, we then allow a time of flight of 10 ms. 

\begin{figure*}
	\centering
		\includegraphics[width=\textwidth]{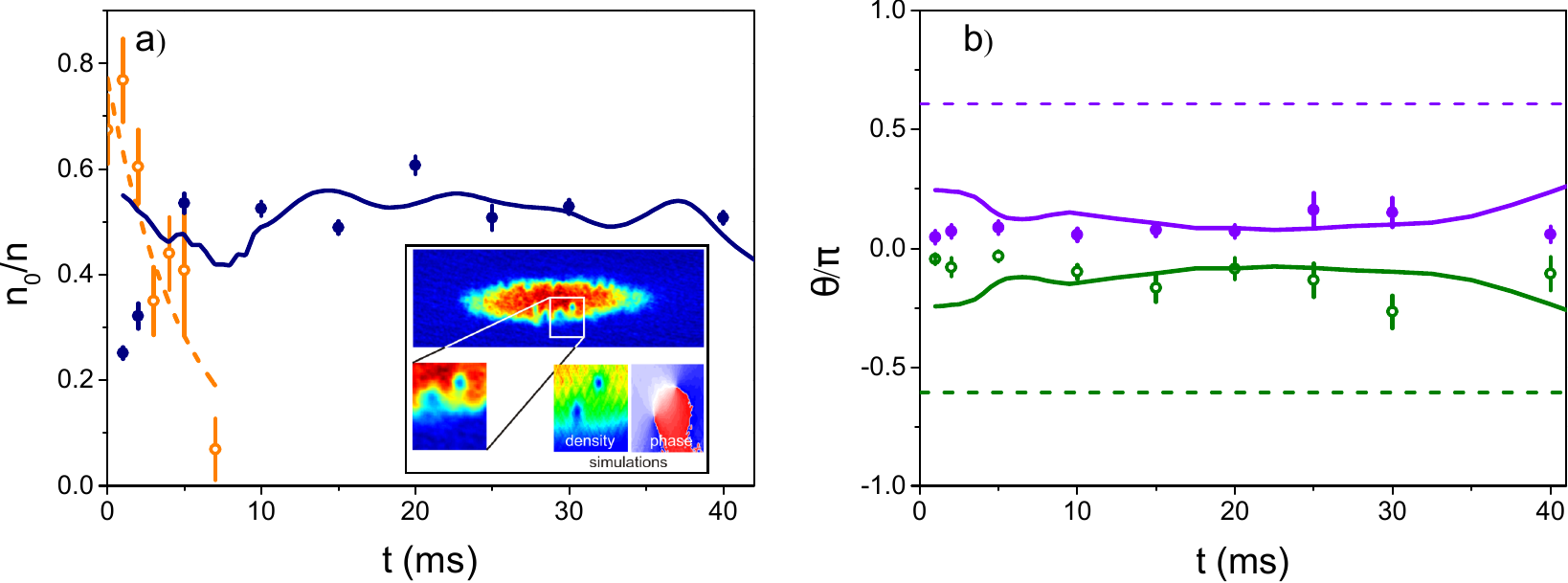}
	\caption{(Color online) a) Points: measured relative depth over time for the JRS (filled blue) and DPS (open orange) as a function of time. Error bars  are one $\sigma$ statistical error. The blue solid line is the depth predicted by the numerical simulations rescaled to take into account the finite resolution of our imaging system ($\simeq$1 $\mu$m). The orange dashed line is an exponential fit to the data. The inset shows an absorption image of our JRSs reaching the border of the BEC and breaking into VA pairs. The experimental observations are in good agreement with the simulations. The phase structure illustrates that the phase winds in opposite directions around the two winding points. b) Points: measured orientation of the JRSs. Error bars are one $\sigma$ statistical error. Filled violet and open green points correspond to solitons travelling towards left and right respectively. The solid lines are the prediction from the numerical simulations. The orientation shows small variations over time due to sound waves travelling through the BEC but a large discrepancy with respect to the direction of travel, indicated by the dashed lines.}
\label{Iw}
\end{figure*}

The experimental evolution that follows the phase imprinting is displayed in figure 2a. In accordance with our simulations (Fig 2b), we observe that the imprinting generates two elliptical rarefaction pulses, one on each side of the triangle. To characterize the motion and the properties of the two travelling JRSs we perform two independent Gaussian fits determining their angle $\theta$, their position, their depth $n_0$, their major ($\alpha$) and minor ($\beta$) axis lengths. We observe that, after the first 5 ms in which the initial sharp triangular imprint decays into the two JRSs, these latter notably travel through the BEC without any form of decay for at least 40ms. At this time, they stop since they reach the border of the BEC. Indeed, as reported in Fig 3a, we observe that until that time, their relative depth $n_0/n$, where $n$ is the density of the unperturbed BEC, remains approximately constant for the whole evolution. The absence of decay as well as the overall dynamical evolution are in agreement with the GPE simulations demonstrating that JRSs are stable solutions of the non-linear Schr\"odinger equation.

To provide a direct comparison, we also study the evolution of a standard DPS created in our experiment imprinting a linear phase step of $\pi$ on the BEC. After only $\simeq$10ms we observe the DPS decaying due to snaking and thermodynamic instability into a pair of vortices. The corresponding depth as a function of time is also reported in Fig 3a. The lifetime obtained by an exponential fit is 4ms \footnote{Our zero-temperature GPE numerical simulations show the onset of the dynamical instability (bending and breaking of the soliton) on the same timescale of the experiment, however we cannot reproduce the fast decay of the amplitude that we observe in the experiment. We therefore attribute this to finite temperature effects, i.e., to the thermodynamic instability. Further investigations on this effect are in order. }, therefore an order of magnitude lower than the lifetime of our JRSs. This latter is limited only by the finite extension of our BEC. 

Interestingly, when reaching the border of the BEC, each JRS breaks into a VA pair \cite{Smirnov,Mironov}, as shown in the inset of Fig 3a. This is due to the fact that at this point the speed of the JRSs rapidly drop to zero making the rarefaction pulses transition to separated VA pairs, as predicted by Jones and Roberts \cite{JR}. This observation, in agreement with our simulations, further confirms that our solitons belong to the Jones-Roberts class. It is worth noticing that the high trapping frequency in the vertical direction prevents the formation of vortex rings. Indeed the smallest vortex ring has a size comparable to four times the healing length $\xi=(8\pi na_s)^{-1/2}$, being $a_s$ the s-wave scattering length and $n$ the peak density. For our BEC the healing length is $\simeq$420 nm while the Thomas-Fermi radius in the vertical direction is 1.1 $\mu$m. Therefore, our condensate cannot support the formation of vortex rings but only of VA pairs along the compressed vertical direction. From this we conclude that even though our BEC is not strictly two-dimensional, it can only support 2D JRSs.

By measuring the speed of propagation we confirm the subsonic nature of our JRSs, as they move with an average speed of 0.43 mm/s, which is smaller than the speed of sound of 1.21 mm/s. A single JRS is expected to travel at an angle  $\theta=$90$^\circ$ with respect to its major axis \cite{JR}. Interestingly the direction of travel of our two JRSs features a slightly different angle, as reported in Fig. 3b. The two JRS tend to arrange in a more (anti)parallel configuration with respect to the initial one suggesting that they might feature a dipolar-like interaction, stemming from their elongated dipolar phase structure.

By analysing the dynamics of their size, we can gain further insights on our JRSs. As reported in Fig. 4, after an initial time of 5 ms in which the initial triangular structure decays into the two JRSs, their major axes $\alpha$ reaches a stable value that is kept until the solitons reach the border of the BEC. This behaviour coincides with the predictions of our numerical simulations (solid line in Fig. 4a). The size of the minor axis $\beta$ is at the limit of our resolution also after 10ms of expansion and we do not observe any significant change, as also  expected from the simulations (Fig. 4b). As can be seen in Fig. 4, once they are formed, our JRSs acquire a shape that fulfils the Kadomtsev-Petviashvili condition \cite{KP} $\alpha=C/\xi^2$ and $\beta=C/\xi$ (dotted lines), with $\xi=\sqrt{1-v/c}/3$  and $C\simeq\xi/3$. The fulfilment of the KPC is another characteristic feature of two-dimensional JRSs with finite velocity \cite{Tsuchi,JR2}. Intuitively an increasing axis length over time in both directions due to dissipation would be expected, similar to DPS becoming wider and faster due to thermodynamic dissipation, as observed in previous experiments \cite{Burger}. Notably, as far as our experiment can test, the fulfilment of the KPC provides an outstanding immunity against both the snaking and the thermodynamic instability, making the scattering of sound waves also negligible \cite{Mironov}.  

\begin{figure}
	\centering
		\includegraphics[width=0.5\textwidth]{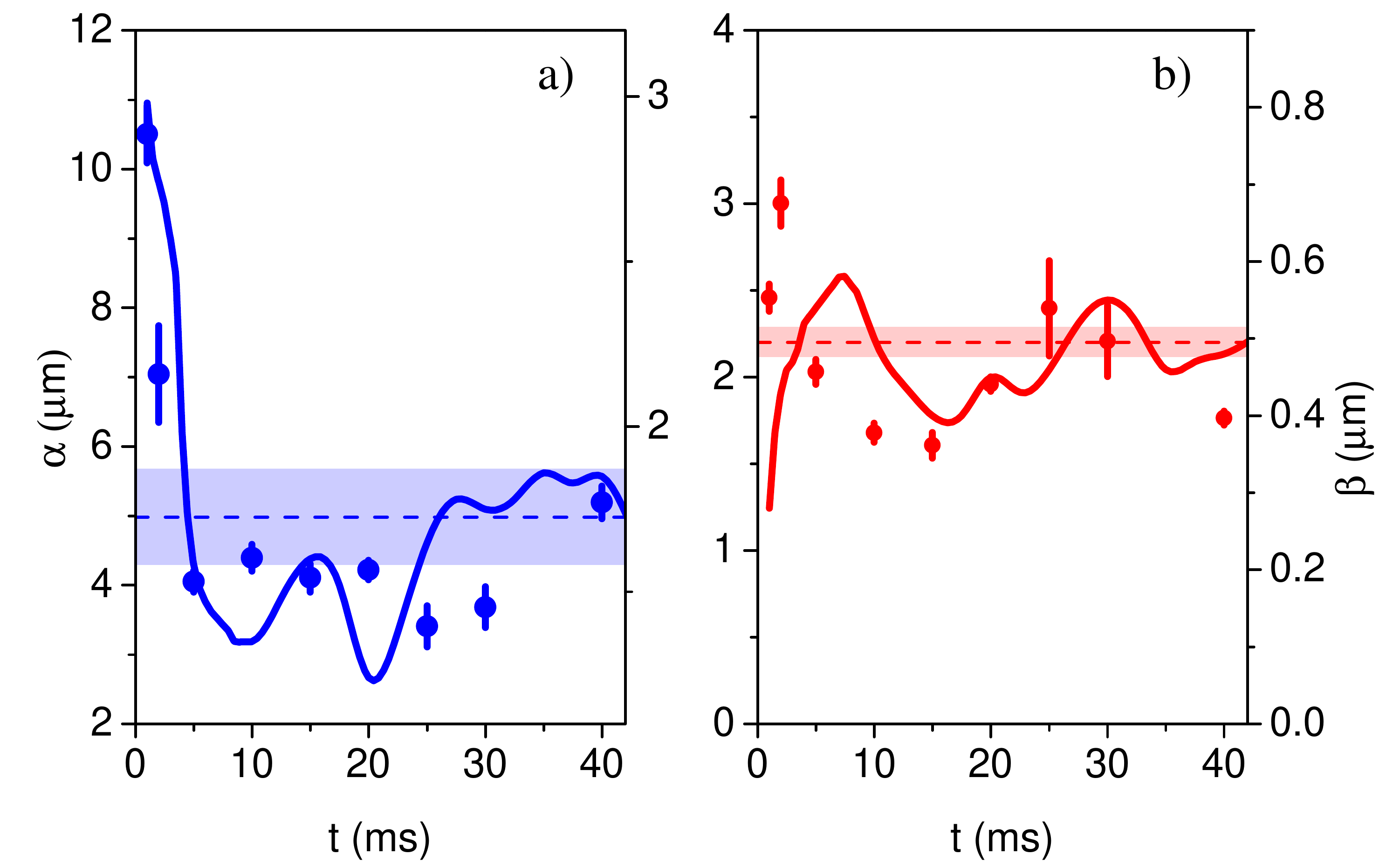}
	\caption{(Color online) a) Evolution of the long axis $\alpha$ of the JRSs as a function of time. Points are the experimental data while the solid line is the result of the numerical simulations. The small wiggles are due to sound waves travelling across the BEC. The scale on the left (right) relates to the experimental data (simulations). The scaling factor of ~3 comes from the 10 ms time of flight and is in accordance with the numerical simulations. The dashed line (related to the scale on the right) is the mean value of the KPC for times $>5$ ms and the shaded band takes into account the fluctuations of the speed of the solitons. b) The same as a) but for the short axis $\beta$. The different scaling factor takes into account also the limited resolution of our imaging ($\simeq$1 $\mu$m). Error bars for a) and b) are one $\sigma$ statistical error. }
\label{Iw}
\end{figure}

In summary, we have experimentally realized and characterized JRSs and confirmed that they are stable solutions of the non-linear Schr\"odinger equation, a long sought goal since their prediction in 1982 \cite{JR,JR2}. By studying their motion and shape, we have found an interesting indication of possible dipole-like interaction and confirmed the fulfilment of the KPC. All this creates an experimental opportunity to investigate the contribution of JRS to the specific heat of the BEC that possibly exceeds the phonon contribution \cite{Tsuchi2}. Furthermore, studying the onset of vortex-free rarefaction pulses can also shed light on the anomalous critical scaling in acoustic turbulences described by the Kardar-Parisi-Zhang equation \cite{Mat}, which is relevant in turbulent systems far from equilibrium such as avalanches \cite{Ager}, formation of fire fronts \cite{Maunu} and surface growth \cite{Arao}. The outstanding resilience of JRSs against dynamical instabilities and thermal decay might allow their propagation in disordered media, suggesting that they can play a significant role in many areas of science. This encourages the search for similar phenomena in other areas of physics, chemistry and biology and opens up novel technology opportunities in directed transport through homogeneous but non-linear media.

\paragraph*{Acknowledgements}
We gratefully thank J. Brand and T. Gasenzer for discussions. This work was funded by the EPSRC (EP/H009914/1) and the Leverhulme trust.


\begin{thebibliography}{}

\bibitem{SR}
J. Scott-Russell. Report on Waves, Proc. Roy. Soc. Edinburgh 319, (1844)

\bibitem{Kiv}
Y. S. Kivshar, and B. Luther-Davies, Dark optical solitons: physics and applications, Phys. Rep. 298, 81 (1998)

\bibitem{Kiv2}
Y. S. Kivshar, and G. Agrawal, Optical Solitons: From Fibers to Photonic Crystals, Academic Press (2003).

\bibitem{Lautrup}
B. Lautrup, R. Appali, A. D. Jackson, and T. Heimburg, The stability of solitons in biomembranes and nerves, Eur. Phys. J. E. Soft Matter 34, 1 (2011)

\bibitem{Heimburg}
T. Heimburg, and A. D. Jackson, On soliton propagation in biomembranes and nerves, PNAS 2005, 9 (2005)

\bibitem{Christie}
D. R. Christie, K. J. Muirhead, and A. L. Hales, Intrusive density flows in the lower troposphere: A source of atmospheric solitons, J. Geophys. Res. 84, 4959 (1979)

\bibitem{Anderson}
B. P. Anderson, et al, Watching Dark Solitons Decay into Vortex Rings in a Bose-Einstein Condensate, Phys. Rev. Lett. 86, 2926 (2001)

\bibitem{Becker}
C. Becker et al., Oscillations and interactions of dark and dark-bright solitons in Bose-Einstein condensates, Nat. Phys. 4, 496 (2008)

\bibitem{Burger}
S. Burger, K. Bongs, S. Dettmer, W. Ertmer, K. Sengstock, A. Sanpera, G. V. Shlyapnikov, and M. Lewenstein, Dark Solitons in Bose-Einstein Condensates, Phys. Rev. Lett. 83,  5198 (1999)

\bibitem{Dens}
J. Denschlag, et al., Generating Solitons by Phase Engineering of a Bose-Einstein Condensate, Science 287, 97 (2000)

\bibitem{Streck}
K. E. Strecker, G. B. Partridge, A. G. Truscott, and R. G. Hulet, Formation and propagation of matter-wave soliton trains, Nature 417, 150 (2002)

\bibitem{Fed}
P. O. Fedichev, A. E. Muryshev, and G. V. Shlyapnikov,  Dissipative dynamics of a kink state in a Bose-condensed gas, Physcial Rev. A 60, 3220 (1999)

\bibitem{Mury}
A. Muryshev, G. V. Shlyapnikov, W. Ertmer, K. Sengstock, M. and Lewenstein, Dynamics of Dark Solitons in Elongated Bose-Einstein Condensates, Phys. Rev. Lett. 89, 110401 (2002).

\bibitem{JR}
C. Jones, and P. H. Roberts, Motions in a Bose condensate. IV. Axisymmetric solitary waves, J. Phys. A. Math. Gen. 15, 2599 (1982)

\bibitem{Smirnov}
L. A. Smirnov, and V. A. Mironov, Dynamics of two-dimensional dark quasisolitons in a smoothly inhomogeneous Bose-Einstein condensate. Phys. Rev. A 85, 053620 (2012)

\bibitem{Tsuchi}
S. Tsuchiya, F. Dalfovo, and L. Pitaevskii, Solitons in two-dimensional Bose-Einstein condensates, Phys. Rev. A 77, 045601 (2008)

\bibitem{Dobrek}
L. Dobrek, et al. Optical generation of vortices in trapped Bose-Einstein condensates, Phys. Rev. A 60, 3381 (1999)

\bibitem{Mironov}
V. A. Mironov, and L. A. Smirnov, Scattering of two-dimensional dark quasi-solitons by smooth inhomogeneities in a Bose-Einstein condensate, Phys. Wave Phenom. 21, 62 (2013) 

\bibitem{KP}
B. B. Kadomtsev, and V. I. Petviashvili, On the stability of solitary waves in weakly dispersing media, Soviet Physics Doklady 15, 539 (1970)

\bibitem{JR2} 
C. Jones, and P.H. Roberts, Motions in a Bose condensate, V. Stability of solitary wavesolutions of non-linear Schrodinger equations in two and three dimensions, J. Phys. A. Math. Gen. 19, 2991 (1986)

\bibitem{Tsuchi2}
S.	Tsuchiya, F. Dalfovo, C. Tozzo, and L. Pitaevskii, Stability and excitations of solitons in 2D Bose-Einstein condensate, J. Low Temp. Phys. 148, 393 (2007)

\bibitem{Mat}
S.	Mathey, T. Gasenzer, and J. M. Pawlowski, Anomalous scaling at nonthermal fixed points of Burgers' and Gross-Pitaevskii turbulence, Phys. Rev. A 92, 023635 (2015)

\bibitem{Ager}
C. M. Aegerter, R. Günther, and R. J. Wijngaarden, Avalanche dynamics, surface roughening, and self-organized criticality: Experiments on a three-dimensional pile of rice, Phys. Rev. E. Stat. Nonlin. Soft Matter Phys. 67, 051306 (2003)

\bibitem{Maunu}
J. Maunuksela, M. Myllys, J. Timonen, M. J. Alalva, and T. Ala-Nissila, Kardar - Parisi - Zhang scaling in kinetic roughening of fire fronts, Physica A 266, 372 (1999)

\bibitem{Arao}
F. D. Aar\~ao Reis, Universality in two-dimensional Kardar-Parisi-Zhang growth. Phys. Rev. E 69, 021610 (2004)

\end{thebibliography}
\end{document}